\documentclass[fleqn,usenatbib]{mnras}

\pdfoutput=1 %for arxiv \pdfoutput=1 within the first 5 lines of the preamble of the main .tex file.
\usepackage{newtxtext,newtxmath}
\usepackage[T1]{fontenc}
\usepackage{ae,aecompl}
\usepackage{hyperref} %To make the eprint and doi bibtex fields hyperlinks
\usepackage{verbatim} %for multiple line comments with \begin{comment}
\usepackage{graphicx}
\usepackage{subcaption}
\usepackage[table,usenames,dvipsnames]{xcolor}

\def\be{\begin{equation}} 
\def\ee{\end{equation}} 
\def\bea{\begin{eqnarray}} 
\def\eea{\end{eqnarray}}

\title[Global 21 cm Signal of a String Wake Network]{The Global 21 cm Signal of a Network of Cosmic String Wakes}{}

\author[O. F. Hern\'andez]{
Oscar F. Hern\'andez$^{1,2}$\thanks{Email: oscarh@physics.mcgill.ca}
\\
$^{1}$Department of Physics, McGill University, 3600 rue University, Montr\'eal, QC, H3A 2T8, Canada
\\
$^{2}$Marianopolis College,  4873 Westmount Ave.,Westmount, QC H3Y 1X9, Canada
}

\date{}
\pubyear{2021}

\begin{document}
\label{firstpage}
\pagerange{\pageref{firstpage}--\pageref{lastpage}}
\maketitle

\begin{abstract}
In previous works we discussed the 21 cm signature of a single cosmic string wake. However the 21 cm brightness temperature is influenced by a network of cosmic string wakes, and not one single wake. In this work we consider the signal from a network of wakes laid down during the matter era. We also improve on the previous calculation of a single wake signature.  Finally we calculate the enhancement of the global 21 cm brightness temperature due to a network of wakes and discuss its affects of the signal measured in the Wouthuysen-Field absorption trough. We estimated that for string tensions between $10^{-8}$ to $10^{-7}$ there would be between a 10\% to a factor 2 enhancement in the signal. 
\end{abstract}

\begin{keywords}
Cosmology -- cosmic background radiation -- dark ages, reionization, first stars  -- diffuse radiation -- intergalactic medium -- cosmology: theory
\end{keywords}

%%%%%%%section divider %%%%%%%%%%%%%%%%%%%

\section{Introduction}
\label{sec:intro}
In~\cite{Brandenberger:2010hi,Hernandez:2012gz,Hernandez:2014cu} we discussed the signature of a single cosmic string wake in a 21 cm intensity map. However in models which lead to cosmic strings, a network of strings will inevitably form in a phase transition in the early universe.  Here we consider the effect that a network of cosmic string wakes will have on the global 21 cm signal. We are particularly interested in the effect these wakes will have on the Wouthuysen-Field (WF) absorption trough in the global signal. 

The Experiment to Detect the Global EoR Signature (EDGES) has reported a detection of a stronger-than-expected absorption feature in the  21 cm spectrum~\citep{Bowman:2018bs}. This absorption feature would occur in the following way.  Before the first luminous sources produced a large enough number of ultraviolet (UV) photons, the 21 cm spin temperature $T_S$ of the cosmic gas was determined by a competition between Compton scattering and collisions. Compton scattering couples $T_S$ to the CMB radiation temperature $T_\gamma$, whereas collisions couple $T_S$ to the much cooler kinetic temperature $T_K$ of the cosmic gas. In the presence of Lyman-$\alpha$ (Ly$\alpha$) UV radiation, hydrogen atoms can change hyperfine state, and hence $T_S$, through the absorption and re-emission of Ly$\alpha$ photons. This is the Wouthuysen-Field (WF) effect~\citep{Wouthuysen:1952hd,Field:1958}.  The UV photons produced by the first galaxies couple $T_S$ to $T_K$ leading to a more negative brightness temperature. Galaxies also produce X-rays which heat the cosmic gas, and eventually reionization begins. While it is possible that the intergalactic medium (IGM) can be heated to the radiation temperature by X-rays before the spin temperature couples to it, generically there will be a WF absorption trough right after cosmic dawn~\citep{Pritchard:2010dc, Mesinger:2013gs, Mirocha:2016fc, Fialkov:2016fm, Park:2019go}. 

Seven years ago we noted that if a WF absorption trough existed, it would lead to the strongest signal from a cosmic string wake~\citep{Hernandez:2014cu}.  If cosmic strings were to exist, the global 21 cm brightness temperature measured by EDGES would be effected not just by one single wake, but by the network of cosmic string wakes. In this paper we recalculate the global 21 cm brightness temperature signal originating from a single wake as well as a network of wakes laid down during the matter era. We begin by reviewing the current experimental limits on the cosmic string tension, and the methods used to obtain them in section~\ref{sec:current-limits}. In section~\ref{sec:wake-network} we describe a cosmic string wake and the string wake network statistics relevant to our calculation. Next, in section~\ref{sec:wake-signal} we discuss the 21 cm signal from one of these wakes and discuss how the optically thin approximation breaks down in certain situations. We take into account that the wake is sometimes an optically thick medium when averaging over wake orientations, something that was not done in previous calculations~\citep{Brandenberger:2010hi,Hernandez:2012gz,Hernandez:2014cu}. In section~\ref{sec:global-signal} we derive the enhancement global signal resulting from a network of cosmic string wakes laid down during the matter era. We focus our attention on the signal between redshifts $z=14$ to $z=30$ which includes the redshift range where the WF absorption trough exists.  Finally in section~\ref{sec:results-conclusions} we discuss our results and present our conclusions.

%%%%%%%section divider %%%%%%%%%%%%%%%%%%%

\section{A Review of Current Limits on the Cosmic String Tension}
\label{sec:current-limits}

Cosmic strings are linear topological defects, remnants of a high-energy phase transition in the very early Universe, that can form in a large class of extensions of the Standard Model.  Their gravitational effects can be parametrized by their string tension $G\mu$, a dimensionless constant where $G$ is Newton's gravitational constant, and $\mu$ is the energy per unit length of the string. Since $\mu$ is proportional to the square of the energy scale of the phase transition, placing upper bounds on the string tension is probing particle physics from the top down in an energy range complementary to that probed by particle accelerators such as the Large Hadron Collider. This string tension is predicted to be between $10^{-8}<G\mu<10^{-6}$ for Grand Unified models, whereas cosmic superstrings have $10^{-12}<G\mu<10^{-6}$~\citep{Copeland:2004dw,Witten:1985tq}. 
Cosmological observations place limits on the string tension with the magnitude of the signal proportional to $G\mu$. 

The gravitational waves emitted by cosmic string loop decay provide a way to detect cosmic strings.  Whereas previous pulsar timing array results~\citep{Arzoumanian:2016dua,Ringeval:2017ew,Arzoumanian:2018} placed upper bounds on the string tension, the most recent NANOGrav 12.5 year data set \citep{Arzoumanian:2020} has found hints of a stochastic gravitational wave background (SGWB) that can be interpreted as coming from cosmic strings. The string tension allowed depends on the cosmic string model. Different Nambu-Goto string models with $10^{-11} \lesssim G\mu \lesssim 10^{-8}$ would explain the SGWB observed by NANOGrav~\citep{Blasi:2021kq,Ellis:2021bq,Bian:2021gk,Samanta:2020tl}. In models of metastable cosmic strings string tensions as large  as $G\mu\sim10^{-7}$ could also explain these results~\citep{Buchmuller:2020ei,Buchmuller:2021ul}. However in Abelian-Higgs cosmic string models the cosmic string loops decay by particle emission versus gravitational waves~\citep{Hindmarsh:2018gi}. Hence if the SGWB reported by NANOGrav is due to Abelian-Higgs cosmic strings, a fraction of those loops are Nambu-Goto-like and survive to radiate gravitationally. \cite{Hindmarsh:2021wg} find that this fraction needs to be between $10^{-3}$ and $0.1$.

The best current limits on the cosmic string tension come from the CMB angular power spectrum.  The Planck collaboration has placed an upper limit on the string tension for Nambu-Goto strings and Abelian-Higgs strings of $G\mu < 1.3 \times 10^{-7}$  and $G\mu < 3.0 \times 10^{-7}$, respectively, at the 95\% CL~\citep{PlanckCollaboration:2014il}.  Finally, there has been much recent research to develop wavelets and machine learning as more sensitive probes of cosmic strings in the CMB~\citep{Hergt:2017dr,McEwen:2017cg,VafaeiSadr:2018hh, VafaeiSadr:2018bc,Ciuca:2017jz,Ciuca:2019hh,Ciuca:2020,Ciuca:2020erratum,Ciuca:2019ww}. If the NANOGrav data were to be the SGWB from strings with $10^{-11} \lesssim G\mu \lesssim 10^{-7}$, the work in~\cite{Ciuca:2020,Ciuca:2020erratum} has show that there is enough information in noisy CMB maps for strings to be detected by machine learning methods. 

%%%%%%%section divider %%%%%%%%%%%%%%%%%%%

\section{The Cosmic String Wake Network}
\label{sec:wake-network}
The cosmic string network consists of long strings with length larger than the Hubble diameter plus a distribution of string loops with radii smaller than the Hubble radius. Both analytical arguments \citep{Hindmarsh:1995bl} and numerical simulations \citep{Allen:1990,Ringeval:2007gf} tell us that the network of cosmic strings will take on a scaling solution in which the average quantities describing the network are invariant in time if measured in Hubble length $H^{-1}(z)$. Thus the string distribution will be statistically independent on time scales larger than the Hubble radius and its time evolution can be characterized by a random walk with a time step comparable to the Hubble radius. At each time step there will be $N_H$ long cosmic strings per Hubble volume.

\cite{Silk:1984ce} pointed out that long strings moving perpendicular to the tangent vector along the string give rise to ``wakes" behind the string in the plane spanned by the tangent vector to the string and the velocity vector. The wake arises as a consequence of the geometry of space behind a long straight string - space perpendicular to the string is conical with a deficit angle given by $\alpha \, = \, 8 \pi G \mu$. From the point of view of an observer travelling with the string it appears that matter streaming by the string obtains a velocity kick of magnitude
$
\delta v \, = \, 4 \pi G \mu \ v_s \gamma_s \, 
$ 
towards the plane behind the string. Here $v_s$ is the velocity of the string,  $\gamma_s$ is the corresponding relativistic gamma factor. This leads to a wedge-shaped region behind the string with twice the background density. 

A cosmic string segment laid down at time $t_i$ will generate a wake whose physical dimensions at that time are
\be \label{eq:initial_size}
  c_1 t_i ~\times~ t_i v_s\gamma_s ~\times~ 4\pi G\mu t_i v_s\gamma_s \, .
\ee
The dimensions $c_1 t_i$ and $t_i v_s\gamma_s$ span the two length dimensions of the wake and are independent of the string tension. They are both of the order of the instantaneous Hubble radius. The third dimension  is the width of the wake, which is much smaller than the length because it is suppressed by the small parameter $G \mu$.

The constant $c_1$ in equation~\ref{eq:initial_size} is the fraction of the long string  along which the transverse velocity is positively correlated. Each long string, spanning a Hubble volume, has approximately $1/c_1$ different segments creating $1/c_1$ different wakes. Thus a scaling solution with $N_H$ long strings per Hubble volume will contribute $N_w = N_H/c_1$ wakes. The value of $c_1$ can be estimated by calculating $\langle v_s(l) v_s(l') \rangle$ with the simulations and code in~\cite{Ringeval:2007gf} and this gives $c_1\approx 0.1$, (C. Ringeval, private communication). 

The three wake dimensions evolve with time.  After being laid down, the lengths Hubble expand, $l(z)=l(z_i) (z_i+1)/(z+1)$, whereas the wake width will grow by gravitational accretion.To analyze the width's growth we use the Zel'dovich approximation~\citep{Zeldovich:1970} for either shock heated~\citep{Brandenberger:2010hi} or a diffuse wakes~\citep{Hernandez:2012gz}.  At a later time, parametrized by redshift $z$, a diffuse wake initialized in the matter era will have grown to physical dimensions~\citep{Hernandez:2014cu}:
\bea
&& l_1(z)\times l_2(z)\times w(z)= \nonumber \\
 && \left(  {{2\over3}c_1\sqrt{z+1\over z_i+1}  \over H(z) }
       \times {{2\over3}v_s\gamma_s\sqrt{z+1\over z_i+1}  \over H(z)} 
       \times  {G\mu~{16\pi\over5}v_s\gamma_s  \sqrt{z_i+1\over z+1} \over H(z) }\right)
\eea
where $z_i$ is the redshift that corresponds to time $t_i$. Note that as time evolves and we move to smaller redshifts, the wake lengths are shrinking in relation to the Hubble radius as $\sqrt(z+1)$ whereas the width is growing by the same factor.  Shock heated wakes will be half as wide.

One of the length dimensions as well as the width depend on the transverse velocity $v_s$ of the long string through the quantity $v_s \gamma_s$.  This can also be extracted from simulations used in~\cite{Ringeval:2007gf} and these give $\langle \gamma_s v_s \rangle \approx 0.4$ (C. Ringeval, private communication), which we will use here, versus the value of $1/\sqrt{3}$ we used in our previous work~\citep{Brandenberger:2010hi,Hernandez:2012gz,Hernandez:2014cu}. %0.4 from 0.3747

The number density of wakes on a sphere at a fixed redshift $z_e$ is also proportional to $\sqrt{v_s \gamma_s}$.  In Appendix~\ref{n2Dderivation} we calculate this as was done in~\cite{Hernandez:2011ima} by modelling the string world sheet as circular with $\pi r_w^2=l_1\times l_2$.  However our result here corrects a few details in the original calculation, such as the erroneous factor of $\cos\theta_s$ in the denominator of equation 3.13 in~\cite{Hernandez:2011ima}.  We find that the two dimensional number density of wakes intersecting  a sphere of radius $R_e=a(z_e)\chi$ is:
\be
n_{2D}(z_i,z_e)= {N_w \over3} \sqrt{\pi c_1 v_s \gamma_s}\,
               H(z_i)^2 \left({z_e+1 \over z_i+1}\right)^2 
               % g(x_i) \sqrt{x_i+1} % radiation-matter factor
               \ ,
\ee
where $N_w$ is the number of strings per Hubble volume. This is the two dimensional number density of wakes that were laid down at a redshift of $z_i$ and observed on a fixed redshift sphere at redshift $z_e$.  

\begin{figure}
\includegraphics[width=0.5\textwidth]{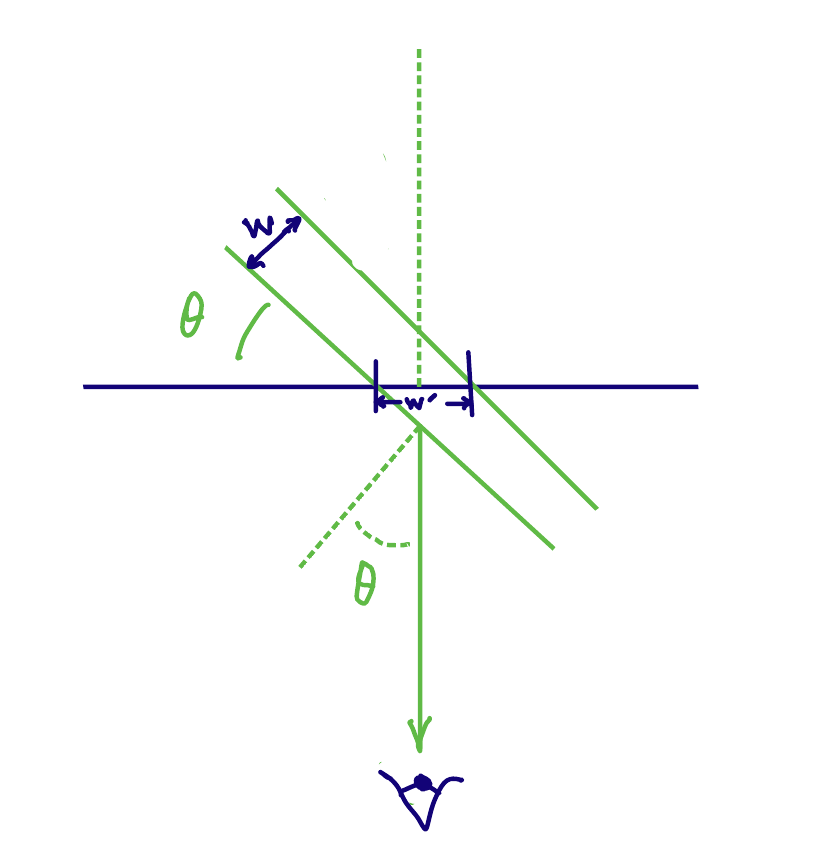}
\caption{\label{fig:wake-intersection}
The wake intersection with a very large sphere with the observer at the centre. $\theta$ is the angle that the normal to the wake makes with the observer. Note that the effective width intersecting the sphere is $w' = w/\sin\theta$.}
\end{figure}

The average area that an intersecting wake covers on the redshift sphere is $ \langle w' \rangle \times 2 r_w $,  where $w$ is the wake width and $w'$ is the effective width intersecting the sphere as shown in figure~\ref{fig:wake-intersection}. Averaging over all solid angle orientations of the wake disc intersecting this sphere we get:
\be
 \langle w' \rangle = \int {d\Omega\over4\pi} {w\over \sin\theta}= {1\over4\pi}\int_0^{\pi} {d\theta} \, \sin\theta  \int_0^{2\pi}  d\phi {w\over \sin\theta}={\pi\over2}w \ .
\ee
Hence the average area  that an intersecting wake covers on the redshift sphere is $ \pi \, w \, r_w $. The fraction of the fixed redshift $z_e$ sphere covered by wakes laid down at redshift $z_i$ is thus
\begin{align}
\label{eq:wake-fraction}
f(z_i,z_e) \, = & \, \pi w \, r_w \, n_{2D}(z_i,z_e)
\nonumber \\
= & \left( {32\pi^2\over45} N_w c_1 v_s^2 \gamma_s^2 \right) G\mu \left({z_e+1\over z_i+1}\right)^2 \, {H_i^2\over H_e^2} 
%\nonumber \\
%& ~~~~~~~~~~~~~~~~~` \sqrt{x_e+1}  \sqrt{x_i+1} [g(x_i)]^2 h(x_i) % radiation-matter factor
\nonumber \\
= & \left( {32\pi^2\over45} N_w c_1 v_s^2 \gamma_s^2 \right) G\mu \,  \left({z_i+1\over z_e+1}\right)
\nonumber \\
\approx &  \, 11 \, G\mu \left({z_i+1\over z_e+1}\right)
\end{align}
In the last line we have used the average values discussed above: $N_w c_1 \approx 10$, $\langle v_s \gamma_s \rangle^2 \approx 0.4^2$.  We now consider the 21 cm radiation from one single wake on this sphere. 

%%%%%%%section divider %%%%%%%%%%%%%%%%%%%

\section{The 21 cm Signal of a Cosmic String Wake}
\label{sec:wake-signal}

From~\cite{Hernandez:2014cu} we have that the optical depth $\tau_\nu$ for a hydrogen cloud or a cosmic string wake is given by
\begin{align}
\label{tau}
 \tau_{\nu}(s)  
 = &  
\frac{3 h c^2 A_{10} }{32 \pi   \nu_{21} k_B } \, { 1 \over T_S}\, [x_{HI} \, n_H \Delta s \, \phi(s,\nu)]
\nonumber \\ 
\approx & \,
[2.583\times 10^{-12}~ {\rm mK cm^2 s^{-1}}]~{ 1 \over T_S}\,  [x_{HI} \, n_H \Delta s \, \phi(s,\nu)]
\end{align}
where $A_{10}=2.85\times10^{-15}\,$s$^{-1}$ is the spontaneous emission coefficient of the 
21 cm transition, $x_{HI}$ is the neutral fraction of hydrogen, $n_H$ is the hydrogen number density, 
$\Delta s$ is the column thickness of the cosmic hydrogen gas or the string wake, $\phi(s,\nu)$ is the 21 cm line profile, and $T_S$ is the spin temperature.

The distinguishing feature between a cosmic gas and a cosmic string wake is the quantity $[x_{HI} \, n_H \Delta s \, \phi(s,\nu)]$.  In particular the column length line profile combination for each is 
\begin{align}
[\Delta s \, \phi(s,\nu)]_{\rm cg} = & \left( {\nu_{21} \over 1+z} H(z) \right)^{-1}
\\
[\Delta s \, \phi(s,\nu)]_{\rm w}            = & \  [\Delta s \, \phi(s,\nu)]_{\rm cg} \ \sin^{-2}(\theta)
\end{align}
for the cosmic gas (cg) and the wake (w), respectively. Here $\theta$ is the angle of the 21~cm ray with respect 
to the vertical to the wake. The optical depth in the cosmic gas is thus,
\begin{align}%{multline}
\tau_{\nu}^{cg}(z) 
= ~&
{[9.075 \times10^{-3}~{\rm K}] \over T_S} ~(z+1)^{3/2}
\nonumber \\ 
& \left({\Omega_b\over0.05}\sqrt{0.3\over\Omega_m}{h\over0.7} \right)
{ x_{HI}(1+\delta_b ) \over \left(1+{\partial{v_{pec}}/\partial{r}\over H(z)/(z+1)}\right) }\ ,
\end{align}%{multline}
where $T_S$ is measured in Kelvin. 

For small string tensions ($G\mu<10^{-7}$) wakes will generically form with no shock heating~\citep{Brandenberger:2010hi,Hernandez:2012gz}.  Hence the wake temperature is not significantly different from that of the IGM and the wake baryon density is twice that of the cosmic gas. Hence the optical depth for the wake is:
\be
\tau_{\nu}^{w}(z) 
\approx  {2\, \tau_{\nu}^{cg}(z) \over \sin^2(\theta)}
\ee
The brightness temperature difference, $\delta T_b(\nu)$ is a comparison of the temperature coming from the hydrogen cloud with the 
``clear view'' of the 21 cm radiation from the CMB~\citep{Furlanetto:2006bq}. 
\be  \label{dTb1}
\delta T_b(\nu) \, = \,
\frac{
T_\gamma(\tau_\nu)-T_\gamma(0)}{1+z} 
= 
\frac{T_S- T_\gamma(0)}{1+z} (1-\exp(-\tau_\nu))
\ee
We usually have an optically thin medium $\tau_\nu<<1$, and hence approximate $(1-\exp(-\tau_\nu))$ by $\tau_\nu$. This approximation holds for the cosmic gas and we will now use it to calculate its brightness temperature. As we will discuss below it does not hold for the brightness temperature of the wake. This point in particular was missed in previous work. 

Observing 21 cm radiation depends crucially on $T_S$. When $T_S$ is above $T_\gamma$ we have emission, when it is below $T_\gamma$ we have absorption. Interaction with CMB photons, spontaneous emission, collisions with hydrogen, electrons, protons, and scattering from UV photons will drive $T_S$ to either $T_\gamma =2.725\, {\rm K} \,(1+z)$ or to $T_K$. Since the times scales for these processes is much smaller than the Hubble time, the spin temperature is determined by  equilibrium in terms of the collision and UV scattering coupling coefficients, $x_c$ and $x_\alpha$, as well as the kinetic and colour temperatures. Before reionization is significant, $x_{HI}\approx1$ and the large optical depth of Ly$\alpha$ photons (given by the Gunn-Peterson optical depth) means that the colour temperature is driven to the kinetic temperature $T_K\approx 0.02\,{\rm K}\,(1+z)^2$, of the IGM.  Thus we can write the spin temperature as~\cite{Furlanetto:2006bq},
\begin{align}\label{eq:spin}
T_{S}\, = ~& T_\gamma \left({ 1+x_c+x_\alpha \over 1+(x_c+x_\alpha)T_\gamma/T_K }\right) 
\\ \nonumber
= ~& 2.725~{\rm K}~ (z+1) \left({ 1+x_c+x_\alpha \over 1+(x_c+x_\alpha)136/(z+1) }\right)  \ . 
\end{align}
If we ignore the peculiar velocities and baryon density fluctuations, and take $\Omega_b=0.05$, $\Omega_m=0.3$, $h=0.7$, the optical depth and brightness temperature for the cosmic gas are
\begin{align} \label{eq:taucg}
\tau_{\nu}^{cg}(z) 
= ~&
3.3303 \times10^{-3}(z+1)^{1/2} \left({ 1+(x_c+x_\alpha)136/(z+1) \over 1+x_c+x_\alpha }\right)
\\ \label{eq:dTbcg}
\delta T_b^{cg}(z) \, \approx &
\, [9.075~{\rm mK}] \, (1+z)^{1/2}
~\left[{x_c+x_{\alpha} \over 1+ x_c+x_{\alpha}}\right]
\left(1-{136\over 1+z}\right)
\ .
\end{align}

\begin{figure}
\centering
\includegraphics[width=0.5\textwidth]{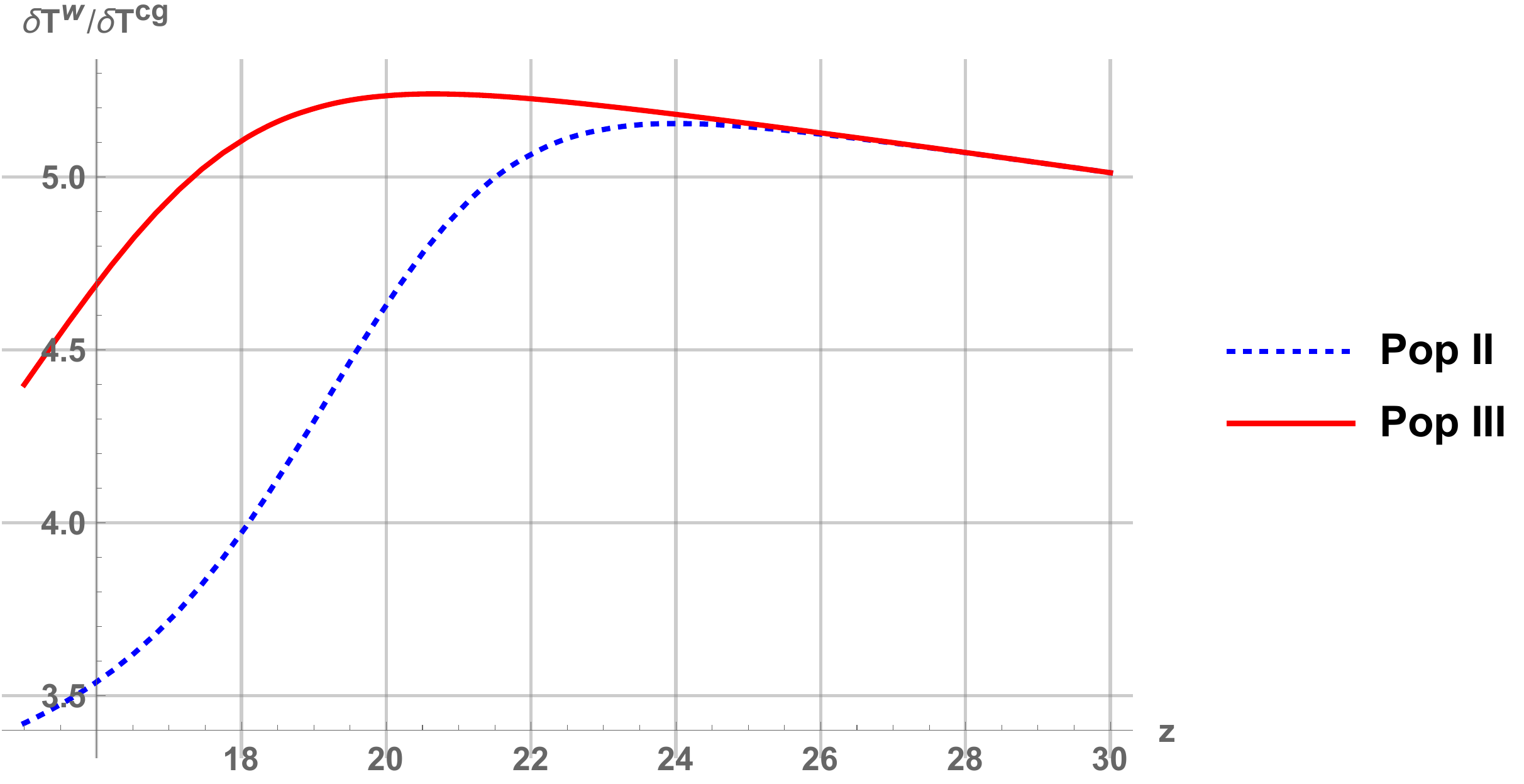}
\caption{\label{fig:dTbratio} The ratio 
%${\delta T_b^{w}(z) \over \delta T_b^{cg}(z)}$
of the average brightness temperature from a cosmic string wake compared to that of the cosmic gas,
for Pop II and Pop III stars.}
\end{figure}

We now consider the brightness temperature in the wake. The $\sin^{-2}$ factor present in cosmic string wakes means that when $\theta$ is near zero the wake optical depth is large. In previous work~\citep{Brandenberger:2010hi,Hernandez:2012gz,Hernandez:2014cu} we erroneously approximated this factor by using $\langle\sin^{2}(\theta)\rangle=1/2$ and the optically thin approximation. However $\langle\sin^{-2}(\theta)\rangle$ diverges and we need to consider the full $(1-\exp(-\tau_\nu))$ factor for the brightness temperature of wakes. We note that the average is taken over all solid angles $d\Omega=d(\cos\theta)d\phi$ and not just $d\theta$. Thus we have that the average brightness temperature from a wake is 
\begin{align} 
\delta T_b^{w}(z) \, \approx &
\frac{T_S- T_\gamma(0)}{1+z} \langle 1-\exp(-\tau_{\nu}^{w}) \rangle 
\\ \nonumber
= & \frac{T_S- T_\gamma(0)}{1+z} 
 \int {d\Omega\over4\pi} ~ \left(1-\exp(- {2\, \tau_{\nu}^{cg} \over \sin^2(\theta)})\right)
\\ \nonumber
= & \frac{T_S- T_\gamma(0)}{1+z} 
 \int_0^{\pi/2} {d\theta} \, \sin\theta \, \left(1-\exp(- {2\, \tau_{\nu}^{cg} \over \sin^2(\theta)})\right)
\end{align}
The $d\theta$ integral can be evaluated in terms of  a Meijer G-function which we Taylor expand since the optical depth of the cosmic gas is small ($\tau_{\nu}^{cg} \lessapprox 0.1$):
\begin{align}
\int_0^{\pi/2} {d\theta} & \sin\theta \, \left(1-\exp(- {2\, \tau\over \sin^2(\theta)})\right) 
\\ \nonumber
&=\tau ( -\ln\tau+1.116 ) + O(\tau^2\ln\tau)
\end{align}
and hence the average brightness temperature from a wake is
\be
\delta T_b^{w}(z) \, \approx 
\frac{T_S- T_\gamma(0)}{1+z} \tau_{\nu}^{cg} ( -\ln\tau_{\nu}^{cg}+1.116 )
\ee
and
\be
{\delta T_b^{w}(z) \over \delta T_b^{cg}(z)} \, \approx  ( -\ln\tau_{\nu}^{cg}+1.116 )
\ee
where $\tau_{\nu}^{cg}$ is given by equation~\ref{eq:taucg}.
Using the analysis of the coefficients $x_c$ and $x_\alpha$ from~\cite{Hernandez:2014cu} we calculate the ratio.  For redshifts below 20 the ratio is dominated by the $x_\alpha$ coefficients and for redshifts above 20 it is dominated by the $x_c$ coefficients. We plot the ratio $\delta T_b^{w}(z) / \delta T_b^{cg}(z)$ in figure~\ref{fig:dTbratio}. We see that at $z=17$ its value is about 3.7 and 4.9 for Pop II and Pop III stars, respectively. 

Eventually x-rays will heat the cosmic gas raising the spin temperature and destroying the WF absorption trough. Our calculation of the brightness temperature does not take this into account, and hence our calculations should not be applied at redshifts below the WF absorption trough, i.e. $z\lessapprox15$. 

%%%%%%%section divider %%%%%%%%%%%%%%%%%%%

\section{The Global 21 cm Signal of the Wake Network}
\label{sec:global-signal}

The fraction $f(z_i,z_e)$ of the fixed redshift $z_e$ sphere covered by wakes that were laid down at redshift $z_i$ is given by equation~\ref{eq:wake-fraction}. The global 21 cm signal from this network of wakes at a redshift $z_e$ that were laid down at a redshift $z_i$ is $ f(z_i,z_e) \times \delta T_b^{w}(z_e)$.  
Since $f(z_i,z_e)$ is proportional to the string tension, it is much less than one the fraction of the sphere covered by wakes laid down during the matter era can be approximated by integrating $f(z_i,z_e)$ between $z_e$ and $z_i=z_{\rm eq}\approx 3400$.  For this result to be accurate to within say 15\%, we would need to check that this integral is less than $\sqrt{0.15}=0.39$. We will see below that for the cases of interest to us this is indeed the case. 

Given the above, the global brightness temperature coming from $z_e$ is approximately
\begin{align} \label{dTbnetworkF}
~&~\delta T_b^{cg}(z_e) \left(1-\int_{z_{\rm eq}}^{z_e} dz_i \, f(z_i,z_e) \right)+
 \delta T_b^{w}(z_e)  \left( \int_{z_{\rm eq}}^{z_e} dz_i \, f(z_i,z_e)  \right)
\nonumber \\ 
~&~\delta T_b^{cg}(z_e)+
\int_{z_{\rm eq}}^{z_e} dz_i \, f(z_i,z_e)  \left( \delta T_b^{w}(z_e)  -   \delta T_b^{cg}(z_e) \right)
\nonumber \\ 
= ~& \delta T_b^{cg}(z_e)
\left[ 1+ \left( 
{\delta T_b^{w}(z_e) \over \delta T_b^{cg}(z_e)} - 1 
\right)
\int_{z_{e}}^{z_{\rm eq}} dz_i \, f(z_i,z_e)   \right]
\end{align}
The quantity in brackets is the factor by which the brightness temperature is enhanced because of the cosmic string wake network. We will call this factor $F$. 
\be
\label{eq:F}
F\equiv \left[ 1+ \left( 
{\delta T_b^{w}(z_e) \over \delta T_b^{cg}(z_e)} - 1 
\right)
\int_{z_{e}}^{z_{\rm eq}} dz_i \, f(z_i,z_e)   \right]
\ee
Using equation~\ref{eq:wake-fraction} we can evaluate the integral in the second term of $F$ with $z_{\rm eq}=3400$:
\be
\int_{z_{e}}^{z_{\rm eq}} dz_i \, f(z_i,z_e)  = 5.5 \, G\mu {(z_{\rm eq} + 1)^2 \over (z_e +1)}={6.5 \times 10^{7}   \over (z_e +1)} G\mu \ . 
\ee
For $z_e > 17$ and $G\mu<10^{-7}$, this integral is less than $0.37$. 

\begin{figure}
\centering
\includegraphics[width=0.5\textwidth]{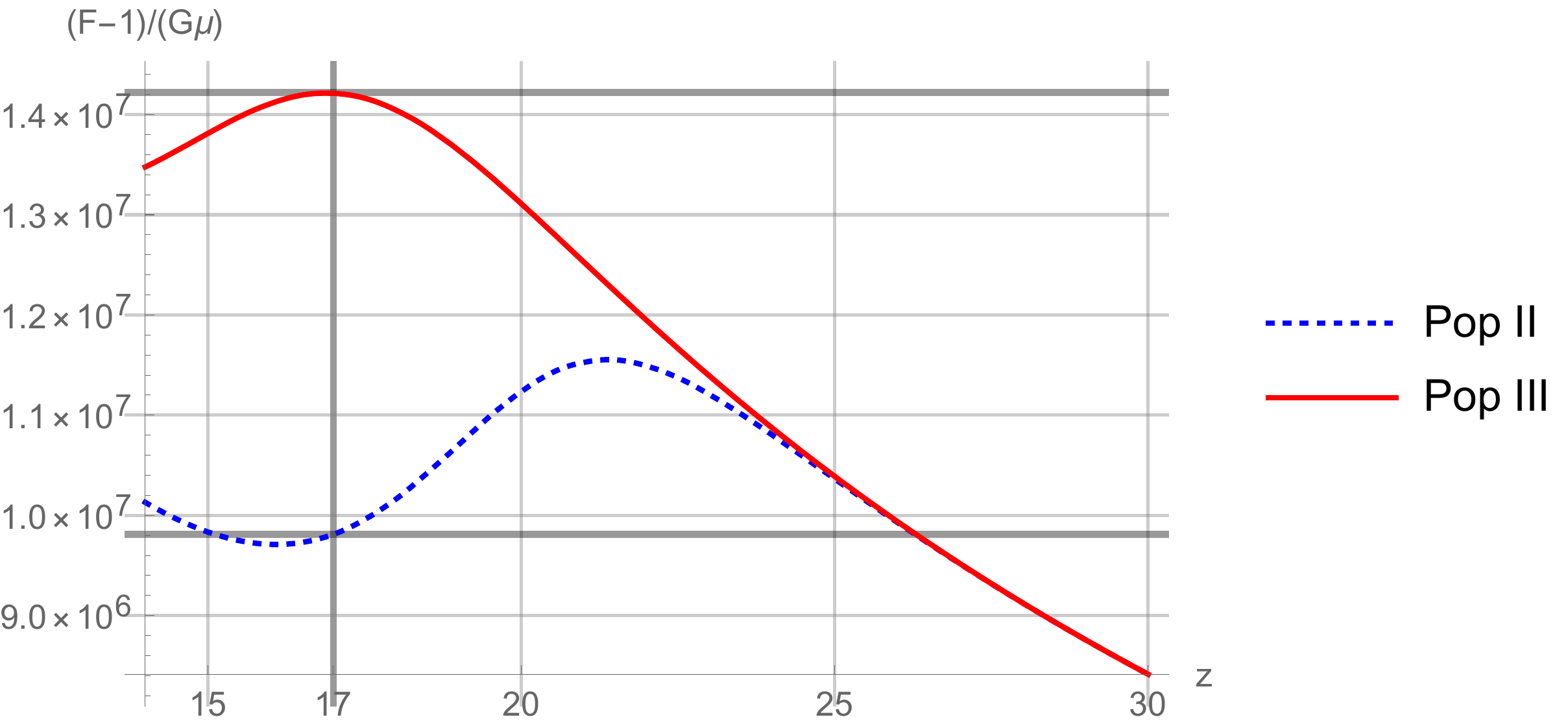}
\caption{\label{fig:Fminus1} The enhancement factor $(F-1)/(G\mu)$ for Pop II and Pop III stars.}
\end{figure}

Thus for wakes laid down during the matter dominated period, we have
\be
F \approx \, 1+ \left( {\delta T_b^{w}(z_e) \over \delta T_b^{cg}(z_e)} - 1 \right) {6.5 \times 10^{7}   \over (z_e +1)} \, G\mu
\ee
In figure~\ref{fig:Fminus1} we plot $(F-1)/(G\mu)$.
At $z_e=17$, which is where the EDGES absorption profile is centred, $F \approx 1+ 0.98 \times 10^{7} ~ G\mu$ and $F \approx 1+ 1.4 \times 10^{7} ~ G\mu$, for Pop II and Pop III stars, respectively. Thus a string tension between $10^{-8}$ to $10^{-7}$ would lead to a 10\% to a factor 2 enhancement in the signal. 

%%%%%%%section divider %%%%%%%%%%%%%%%%%%%

\section{Discussion and Conclusions}
\label{sec:results-conclusions}

The work in~\cite{Hernandez:2014cu} calculated the global 21 cm signal of a single cosmic string wake and made the case that such a place is the best place to look for them. In the present work we have improved on the analysis by improving the calculation of the signal of one wake and considering the effect that a network of wakes laid down during the matter era would have on the signal. We estimated that for string tensions between $10^{-8}$ to $10^{-7}$ there would be between a 10\% to a factor 2 enhancement in the signal. 

The most recent NANOGrav results on the stochastic gravitational wave background admit a cosmic string interpretation with string tensions between $10^{-11}$ to $10^{-7}$.  The encouraging results of this first analysis justifies a more thorough and rigorous work which would consider the contribution of wakes initialized during the radiation era and numerical calculations of quantities we have estimated (work in progress with C. Ringeval). 

%Comparison of cosmic string and superstring models to NANOGrav 12.5-year results~\cite{BlancoPillado:2021uf}. 

%Constraints on Cosmic Strings Using Data from the Third Advanced LIGO-Virgo Observing Run~\cite{AbbottLIGO:2021}

%%%%%%%section divider %%%%%%%%%%%%%%%%%%%

\section*{Acknowledgements}

I am very grateful to Christophe Ringeval for many fruitful discussions and comments on this work. This work was made possible by the support of the Fonds de recherche du Qu\'ebec -- Nature et technologies (FRQNT) Programme de recherche pour les enseignants de coll\`ege, (funding reference number 2021-CO-283996).

\section*{Data Availability}
The data underlying this article will be shared on reasonable request to the corresponding author.

\bibliographystyle{mnras}
\bibliography{wakes21cm}

\begin{thebibliography}{}
\makeatletter
\relax
\def\mn@urlcharsother{\let\do\@makeother \do\$\do\&\do\#\do\^\do\_\do\%\do\~}
\def\mn@doi{\begingroup\mn@urlcharsother \@ifnextchar [ {\mn@doi@}
  {\mn@doi@[]}}
\def\mn@doi@[#1]#2{\def\@tempa{#1}\ifx\@tempa\@empty \href
  {http://dx.doi.org/#2} {doi:#2}\else \href {http://dx.doi.org/#2} {#1}\fi
  \endgroup}
\def\mn@eprint#1#2{\mn@eprint@#1:#2::\@nil}
\def\mn@eprint@arXiv#1{\href {http://arxiv.org/abs/#1} {{\tt arXiv:#1}}}
\def\mn@eprint@dblp#1{\href {http://dblp.uni-trier.de/rec/bibtex/#1.xml}
  {dblp:#1}}
\def\mn@eprint@#1:#2:#3:#4\@nil{\def\@tempa {#1}\def\@tempb {#2}\def\@tempc
  {#3}\ifx \@tempc \@empty \let \@tempc \@tempb \let \@tempb \@tempa \fi \ifx
  \@tempb \@empty \def\@tempb {arXiv}\fi \@ifundefined
  {mn@eprint@\@tempb}{\@tempb:\@tempc}{\expandafter \expandafter \csname
  mn@eprint@\@tempb\endcsname \expandafter{\@tempc}}}

\bibitem[\protect\citeauthoryear{Allen \& Shellard}{Allen \&
  Shellard}{1990}]{Allen:1990}
Allen B.,  Shellard E. P.~S.,  1990, \mn@doi [Phys. Rev. Lett.]
  {10.1103/PhysRevLett.64.119}, 64, 119

\bibitem[\protect\citeauthoryear{Arzoumanian et~al.,}{Arzoumanian
  et~al.}{2016}]{Arzoumanian:2016dua}
Arzoumanian Z.,  et~al., 2016, \mn@doi [ApJ] {10.3847/0004-637X/821/1/13}, 821,
  13

\bibitem[\protect\citeauthoryear{Arzoumanian et~al.,}{Arzoumanian
  et~al.}{2018}]{Arzoumanian:2018}
Arzoumanian Z.,  et~al., 2018, \mn@doi [ApJ] {10.3847/1538-4357/aabd3b}, 859,
  47

\bibitem[\protect\citeauthoryear{Arzoumanian et~al.,}{Arzoumanian
  et~al.}{2020}]{Arzoumanian:2020}
Arzoumanian Z.,  et~al., 2020, \mn@doi [ApJL] {10.3847/2041-8213/abd401}, 905,
  L34

\bibitem[\protect\citeauthoryear{Bian, Cai, Liu, Yang  \& Zhou}{Bian
  et~al.}{2021}]{Bian:2021gk}
Bian L.,  Cai R.-G.,  Liu J.,  Yang X.-Y.,   Zhou R.,  2021, \mn@doi [Phys.
  Rev. D] {10.1103/PhysRevD.103.L081301}, 103, L081301

\bibitem[\protect\citeauthoryear{Blasi, Brdar  \& Schmitz}{Blasi
  et~al.}{2021}]{Blasi:2021kq}
Blasi S.,  Brdar V.,   Schmitz K.,  2021, \mn@doi [Phys. Rev. Lett.]
  {10.1103/PhysRevLett.126.041305}, 126, 041305

\bibitem[\protect\citeauthoryear{Bowman, Rogers, Monsalve, Mozdzen  \&
  Mahesh}{Bowman et~al.}{2018}]{Bowman:2018bs}
Bowman J.~D.,  Rogers A. E.~E.,  Monsalve R.~A.,  Mozdzen T.~J.,   Mahesh N.,
  2018, \mn@doi [Nature] {10.1038/nature25792}, 555, 67

\bibitem[\protect\citeauthoryear{Brandenberger, Danos, Hern{\'a}ndez  \&
  Holder}{Brandenberger et~al.}{2010}]{Brandenberger:2010hi}
Brandenberger R.~H.,  Danos R.~J.,  Hern{\'a}ndez O.~F.,   Holder G.~P.,  2010,
  \mn@doi [J. Cosmol. Astropart. Phys.] {10.1088/1475-7516/2010/12/028}, 2010,
  028

\bibitem[\protect\citeauthoryear{Buchmuller, Domcke  \& Schmitz}{Buchmuller
  et~al.}{2020}]{Buchmuller:2020ei}
Buchmuller W.,  Domcke V.,   Schmitz K.,  2020, \mn@doi [Phys. Lett. B]
  {10.1016/j.physletb.2020.135914}, 811, 135914

\bibitem[\protect\citeauthoryear{Buchmuller, Domcke  \& Schmitz}{Buchmuller
  et~al.}{2021}]{Buchmuller:2021ul}
Buchmuller W.,  Domcke V.,   Schmitz K.,  2021, ArXiv e-print, 2107.04578

\bibitem[\protect\citeauthoryear{Ciuca \& Hern{\'a}ndez}{Ciuca \&
  Hern{\'a}ndez}{2017}]{Ciuca:2017jz}
Ciuca R.,  Hern{\'a}ndez O.~F.,  2017, \mn@doi [J. Cosmol. Astropart. Phys.]
  {10.1088/1475-7516/2017/08/028}, 2017, 028

\bibitem[\protect\citeauthoryear{Ciuca \& Hern{\'a}ndez}{Ciuca \&
  Hern{\'a}ndez}{2019}]{Ciuca:2019hh}
Ciuca R.,  Hern{\'a}ndez O.~F.,  2019, \mn@doi [MNRAS] {10.1093/mnras/sty3478},
  483, 5179

\bibitem[\protect\citeauthoryear{Ciuca \& Hern{\'a}ndez}{Ciuca \&
  Hern{\'a}ndez}{2020}]{Ciuca:2020}
Ciuca R.,  Hern{\'a}ndez O.~F.,  2020, \mn@doi [MNRAS] {10.1093/mnras/stz3551},
  492, 1329

\bibitem[\protect\citeauthoryear{Ciuca \& Hern{\'a}ndez}{Ciuca \&
  Hern{\'a}ndez}{2021}]{Ciuca:2020erratum}
Ciuca R.,  Hern{\'a}ndez O.~F.,  2021, \mn@doi [MNRAS]
  {10.1093/mnras/stab1636}, 506, 1406

\bibitem[\protect\citeauthoryear{Ciuca, Hern{\'a}ndez  \& Wolman}{Ciuca
  et~al.}{2019}]{Ciuca:2019ww}
Ciuca R.,  Hern{\'a}ndez O.~F.,   Wolman M.,  2019, \mn@doi [MNRAS]
  {10.1093/mnras/stz491}, 485, 1377

\bibitem[\protect\citeauthoryear{Copeland, Myers  \& Polchinski}{Copeland
  et~al.}{2004}]{Copeland:2004dw}
Copeland E.~J.,  Myers R.~C.,   Polchinski J.,  2004, \mn@doi [J. High Energy
  Phys.] {10.1088/1126-6708/2004/06/013}, 2004, 013

\bibitem[\protect\citeauthoryear{Ellis \& Lewicki}{Ellis \&
  Lewicki}{2021}]{Ellis:2021bq}
Ellis J.,  Lewicki M.,  2021, \mn@doi [Phys. Rev. Lett.]
  {10.1103/PhysRevLett.126.041304}, 126, 041304

\bibitem[\protect\citeauthoryear{Fialkov, Cohen, Barkana  \& Silk}{Fialkov
  et~al.}{2016}]{Fialkov:2016fm}
Fialkov A.,  Cohen A.,  Barkana R.,   Silk J.,  2016, \mn@doi [MNRAS]
  {10.1093/mnras/stw2540}, 464, 3498

\bibitem[\protect\citeauthoryear{Field}{Field}{1958}]{Field:1958}
Field G.~B.,  1958, Proceedings of the IRE, 46, 240

\bibitem[\protect\citeauthoryear{Furlanetto, Oh  \& Briggs}{Furlanetto
  et~al.}{2006}]{Furlanetto:2006bq}
Furlanetto S.~R.,  Oh S.~P.,   Briggs F.~H.,  2006, \mn@doi [Physics Reports]
  {10.1016/j.physrep.2006.08.002}, 433, 181

\bibitem[\protect\citeauthoryear{Hergt, Amara, Brandenberger, Kacprzak  \&
  Refregier}{Hergt et~al.}{2017}]{Hergt:2017dr}
Hergt L.,  Amara A.,  Brandenberger R.~H.,  Kacprzak T.,   Refregier A.,  2017,
  \mn@doi [J. Cosmol. Astropart. Phys.] {10.1088/1475-7516/2017/06/004}, 2017,
  004

\bibitem[\protect\citeauthoryear{Hern{\'a}ndez}{Hern{\'a}ndez}{2014}]{Hernandez:2014cu}
Hern{\'a}ndez O.~F.,  2014, \mn@doi [Phys. Rev. D]
  {10.1103/PhysRevD.90.123504}, 90, 123504

\bibitem[\protect\citeauthoryear{Hern{\'a}ndez \& Brandenberger}{Hern{\'a}ndez
  \& Brandenberger}{2012}]{Hernandez:2012gz}
Hern{\'a}ndez O.~F.,  Brandenberger R.~H.,  2012, \mn@doi [J. Cosmol.
  Astropart. Phys.] {10.1088/1475-7516/2012/07/032}, 2012, 032

\bibitem[\protect\citeauthoryear{Hern{\'a}ndez, Wang, Fong  \&
  Brandenberger}{Hern{\'a}ndez et~al.}{2011}]{Hernandez:2011ima}
Hern{\'a}ndez O.~F.,  Wang Y.,  Fong J.,   Brandenberger R.~H.,  2011, \mn@doi
  [J. Cosmol. Astropart. Phys.] {10.1088/1475-7516/2011/08/014}, 2011, 014

\bibitem[\protect\citeauthoryear{Hindmarsh \& Kibble}{Hindmarsh \&
  Kibble}{1995}]{Hindmarsh:1995bl}
Hindmarsh M.,  Kibble T. W.~B.,  1995, \mn@doi [Reports on Progress in Physics]
  {10.1088/0034-4885/58/5/001}, 58, 477

\bibitem[\protect\citeauthoryear{Hindmarsh, Lizarraga, Urrestilla, Daverio  \&
  Kunz}{Hindmarsh et~al.}{2018}]{Hindmarsh:2018gi}
Hindmarsh M.,  Lizarraga J.,  Urrestilla J.,  Daverio D.,   Kunz M.,  2018,
  \mn@doi [Phys. Rev. D] {10.1103/PhysRevD.99.083522}, 99, 449

\bibitem[\protect\citeauthoryear{Hindmarsh, Lizarraga, Urio  \&
  Urrestilla}{Hindmarsh et~al.}{2021}]{Hindmarsh:2021wg}
Hindmarsh M.,  Lizarraga J.,  Urio A.,   Urrestilla J.,  2021, ArXiv e-print,
  2103.16248

\bibitem[\protect\citeauthoryear{McEwen, Feeney, Peiris, Wiaux, Ringeval  \&
  Bouchet}{McEwen et~al.}{2017}]{McEwen:2017cg}
McEwen J.~D.,  Feeney S.~M.,  Peiris H.~V.,  Wiaux Y.,  Ringeval C.,   Bouchet
  F.~R.,  2017, \mn@doi [MNRAS] {10.1093/mnras/stx2268}, 472, 4081

\bibitem[\protect\citeauthoryear{Mesinger, Ferrara  \& Spiegel}{Mesinger
  et~al.}{2013}]{Mesinger:2013gs}
Mesinger A.,  Ferrara A.,   Spiegel D.~S.,  2013, \mn@doi [MNRAS]
  {10.1093/mnras/stt198}, 431, 621

\bibitem[\protect\citeauthoryear{Mirocha, Furlanetto  \& Sun}{Mirocha
  et~al.}{2016}]{Mirocha:2016fc}
Mirocha J.,  Furlanetto S.~R.,   Sun G.,  2016, \mn@doi [MNRAS]
  {10.1093/mnras/stw2412}, 464, 1365

\bibitem[\protect\citeauthoryear{Park, Mesinger, Greig  \& Gillet}{Park
  et~al.}{2019}]{Park:2019go}
Park J.,  Mesinger A.,  Greig B.,   Gillet N.,  2019, \mn@doi [MNRAS]
  {10.1093/mnras/stz032}, 484, 933

\bibitem[\protect\citeauthoryear{{Planck Collaboration} et~al.,}{{Planck
  Collaboration} et~al.}{2014}]{PlanckCollaboration:2014il}
{Planck Collaboration} et~al., 2014, \mn@doi [A{\&}A]
  {10.1051/0004-6361/201321621}, 571, A25

\bibitem[\protect\citeauthoryear{Pritchard \& Loeb}{Pritchard \&
  Loeb}{2010}]{Pritchard:2010dc}
Pritchard J.~R.,  Loeb A.,  2010, \mn@doi [Phys. Rev. D]
  {10.1103/PhysRevD.82.023006}, 82, 023006

\bibitem[\protect\citeauthoryear{Ringeval \& Suyama}{Ringeval \&
  Suyama}{2017}]{Ringeval:2017ew}
Ringeval C.,  Suyama T.,  2017, \mn@doi [J. Cosmol. Astropart. Phys.]
  {10.1088/1475-7516/2017/12/027}, 2017, 027

\bibitem[\protect\citeauthoryear{Ringeval, Sakellariadou  \& Bouchet}{Ringeval
  et~al.}{2007}]{Ringeval:2007gf}
Ringeval C.,  Sakellariadou M.,   Bouchet F.~R.,  2007, \mn@doi [J. Cosmol.
  Astropart. Phys.] {10.1088/1475-7516/2007/02/023}, 2007, 023

\bibitem[\protect\citeauthoryear{Samanta \& Datta}{Samanta \&
  Datta}{2020}]{Samanta:2020tl}
Samanta R.,  Datta S.,  2020, arXiv, 2009.13452v3

\bibitem[\protect\citeauthoryear{Silk \& Vilenkin}{Silk \&
  Vilenkin}{1984}]{Silk:1984ce}
Silk J.,  Vilenkin A.,  1984, \mn@doi [Phys. Rev. Lett.]
  {10.1103/PhysRevLett.53.1700}, 53, 1700

\bibitem[\protect\citeauthoryear{Vafaei~Sadr, Movahed, Farhang, Ringeval,
  Bouchet  \& Bouchet}{Vafaei~Sadr et~al.}{2018a}]{VafaeiSadr:2018hh}
Vafaei~Sadr A.,  Movahed S. M.~S.,  Farhang M.,  Ringeval C.,  Bouchet F.~R.,
  Bouchet F.~R.,  2018a, \mn@doi [MNRAS] {10.1093/mnras/stx3126}, 475, 1010

\bibitem[\protect\citeauthoryear{Vafaei~Sadr, Farhang, Movahed, Bassett  \&
  Kunz}{Vafaei~Sadr et~al.}{2018b}]{VafaeiSadr:2018bc}
Vafaei~Sadr A.,  Farhang M.,  Movahed S. M.~S.,  Bassett B.,   Kunz M.,  2018b,
  \mn@doi [MNRAS] {10.1093/mnras/sty1055}, 478, 1132

\bibitem[\protect\citeauthoryear{Witten}{Witten}{1985}]{Witten:1985tq}
Witten E.,  1985, \mn@doi [Phys. Lett. B] {10.1016/0370-2693(85)90540-4}, 153,
  243

\bibitem[\protect\citeauthoryear{Wouthuysen}{Wouthuysen}{1952}]{Wouthuysen:1952hd}
Wouthuysen S.~A.,  1952, \mn@doi [AJ] {10.1086/106661}, 57, 31

\bibitem[\protect\citeauthoryear{Zel'dovich}{Zel'dovich}{1970}]{Zeldovich:1970}
Zel'dovich Y.~B.,  1970, A{\&}A, 5, 84

\makeatother
\end{thebibliography}

\appendix

\section{The wake number density} 
\label{n2Dderivation}
The two dimensional number density $n_{2D}$ of wakes intersecting a sphere inside a Hubble volume is given by the product of the expected number of string wakes laid down in a Hubble volume, $N_w$, and the probability $P$ of a wake intersecting that sphere, divided by the surface area $4\pi R^2$ of the sphere: $n_{2D}= N_w P/(4\pi R^2)$.  We calculate this below. 

\begin{figure}
\includegraphics[width=0.5\textwidth]{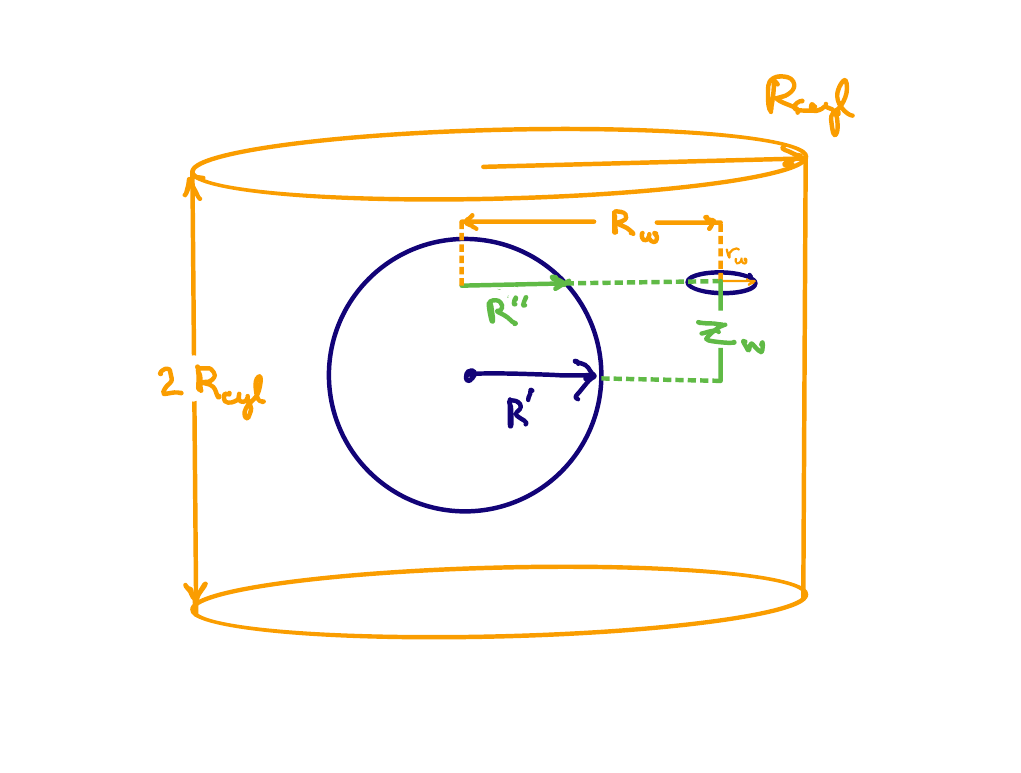}
\caption{\label{fig:cylinder}
A cylindrically shaped Hubble volume containing a string wake and a sphere.}
\end{figure}

We consider a string wake of physical dimensions $l_1\times l_2\times w$ laid down at time $t_i$. To simplify our analysis we model this wake  as an equal volume thin coin-like disc of radius $r_w$, where $\pi r_w^2= l_1 l_2$. Consider a physical Hubble volume at this same time $t_i$ that contains this wake and a sphere of physical radius $R'$. We take the Hubble volume to have the shape of a cylinder with axis parallel to the wake disc axis, as shown in figure~\ref{fig:cylinder}.
The probability $P$ is the volume where the wake disc touches the $R'$ sphere divided by the cylinder's volume. Set up cylindrical coordinates at the origin of this sphere and let $z_w, R_w, \phi_w$ be the coordinates of the wake. Then
\bea
P= {1\over V_{\rm cylinder}}  \int_{-R'}^{R'} dz_w \int_{0}^{R_{\rm cylinder}} dR_w \int_{0}^{2\pi} d\phi_w   \int_{0}^{R_{\rm cylinder}} dR'' &
\nonumber\\ 
\Big[\Theta((R''+r_w)-R_w) \Theta(R_w-(R''-r_w)) ~~~~~~~~~~~&
\nonumber\\ 
\delta^D(R''-(R'^2-z_w^2)^{1/2}) \Big] ~~~~~~~~~&
\nonumber\\ 
={2\pi^2 R'^2 r_w \over V_{\rm cylinder}} ~~~~~~~~~~~~~~~~~~~~~~~~~~~~~~~~~~~~~~~~~~~~~~~~~~~~~~~~~~~~~~~~~~~~~~~~~~&
\eea
Since $V_{\rm cylinder} \approx H(t_i)^3$ we have that the expected number of wakes intersecting the sphere at time $t_i$ is
\be \label{eq:NwP}
\langle N_{S^2} \rangle = N_w P = N_w H_i^3 2\pi^2 R'^2 r_w
\ee

We are interested in the two dimensional number density $n_{2D}$ of wakes that were laid down at a redshift of $z_i$ and observed on a fixed redshift sphere at redshift $z_e$. The sphere from which the observed radiation is emitted has $R_e=a_e \chi_e$, but it had radius $R'=a_i \chi_i$ when the wakes were laid down. And thus dividing equation~\ref{eq:NwP} by $4\pi R_e^2$ we get
\bea
n_{2D}(z_i,z_e) &=& N_w H_i^3 2\pi^2 r_w  \left({z_e+1 \over z_i+1}\right)^2
\\ \nonumber
&=& {N_w \over3} \sqrt{\pi c_1 v_s \gamma_s}\,
               H(z_i)^2 \left({z_e+1 \over z_i+1}\right)^2
\eea
where in the last line we have used $r_w= H_i^{-1} \, 2/3 \sqrt{c_1v_s\gamma_s/\pi} $ as can be seen from equation~\ref{eq:initial_size}. 
 
% Don't change these lines
\bsp	% typesetting comment
\label{lastpage}
\end{document}